\documentclass[12pt]{article}
\usepackage{fullpage}
\usepackage[utf8]{inputenc}
\usepackage{color}
\usepackage{fullpage}
\usepackage{graphicx}
\usepackage{psfrag}
\usepackage{graphicx,psfrag}
\usepackage{subfig}
\usepackage{epsfig}
\usepackage{color}
\usepackage{url}
\usepackage{setspace}
\usepackage{amsmath,amsfonts,amssymb,amsthm}
 \usepackage{ulem}
\usepackage{authblk}
\usepackage[title,titletoc]{appendix}


\providecommand{\keywords}[1]
{
  \small	
  \textbf{\textit{Keywords :}} #1
}
\usepackage{amsfonts,amssymb,amsthm}
\usepackage{amsfonts, amssymb}
\usepackage{hyperref}
\hypersetup{hidelinks}
\usepackage{algorithm}
\usepackage{algorithmicx, algpseudocode}

\makeatletter
\let\@fnsymbol\@arabic
\makeatother

\title{False Discovery Rate Controlling Procedures with  BLOSUM62 substitution matrix  and their application to HIV Data}
\author[]{
Kyurhi Kim \thanks{Department of Statistics, Seoul National University, Gwanak-ro, Gwanak-gu, Seoul 08826, Korea}, 
Junyong Park \thanks{Department of Statistics, Seoul National University, Gwanak-ro, Gwanak-gu, Seoul 08826, Korea, \texttt{junyongpark@snu.ac.kr}},
Dohwan Park \thanks{Department of Mathematics and Statistics, University of Maryland Baltimore County, MD 21250, USA}, 
Mileidy Giraldo, 
\thanks{Amazon Web Services, USA}, 
 Muriel Aldunate 
 \thanks {Centre for Biomedical Research, Burnet Institute, Melbourne, VIC, Australia},  
John L. Spouge \thanks{National Center for Biotechnology Information, National Library of Medicine, National Institutes of Health, Bethesda, Maryland 20894, U.S.A.}, 
Gilda Tachedjian$^5$ 
}
\date{}

\begin{document}

\maketitle
\begin{abstract}
\setstretch{1.5}{
Identifying significant sites in sequence data and analogous data is of fundamental importance in many biological fields. Fisher's exact test is a popular technique, however this approach to sparse count data is not appropriate due to conservative decisions. Since count data in HIV data are typically very sparse, it is crucial to use additional information to statistical models to improve testing power. In order to develop new approaches to incorporate biological information in the false discovery controlling procedure, we 
propose two models: one based on the empirical Bayes model under independence of  amino acids and  the other uses pairwise associations of amino acids based on Markov random field with on the BLOSUM62 substitution matrix. 
We apply the proposed methods to HIV data and  identify significant sites incorporating BLOSUM62 matrix while the traditional method based on Fisher's test 
does not discover any site. 
These newly developed methods have the potential to handle many biological problems in the studies of vaccine and drug trials and phenotype studies. 
\\ \\
}
\end{abstract}


\section{Introduction}
Many biological data sets naturally suggest testing a null hypothesis that two independent sets have been sampled from a single population (“two-independent sample tests”). Although tests based on matched sampling offer attractive properties, obtaining matching variables or collecting matched data is sometimes inappropriate or even impossible. Any experimental setup relying on the random assignment of treatment to members of the same sample or drawing samples at random from the same population is by definition a two-independent sample case. For instance, vaccine and drug trials with their treated vs. placebo/untreated cohorts, studies analyzing disease-susceptible vs. disease-resistant groups, and any differential phenotype analyses: e.g. fast- vs. slow-growing, antibiotic-resistant vs. antibiotic-susceptible, evolutionarily fit vs. unfit groups, etc. With the advent of large-scale genomics and next-generation sequencing, biologists often use sequence data to frame these types of questions in search of  molecular correlates for a trait of interest. A common research strategy is to compare alignments of DNA or protein sequences in the two groups and to perform a statistical test procedure on the distribution of residues (DNA or protein letters) observed in the two groups at each position of the sequence alignment.

In the field of of HIV vaccine design, one main focus is on identifying the properties that distinguish HIV founder viruses from other circulating viruses in the donor. 
In this context, one particular interest is to identify signature residues which are defined as protein sites determining that the amino acids preference in the founder viruses is significantly different from the corresponding residue in the donor viruses. The identification of these signature sites plays a crucial role in understanding the mechanisms underlying the HIV infection. 
For instance, see \cite{Gonzalez2017} and \cite{Gonzalez2015}. 
\cite{Smith} used the Fisher's test in \cite{Fisher} to analyze donor vs. recipient sequences to study transmission of Human Immunodeficiency Virus (HIV) and Simian Immunodeficiency Virus. However this Fisher's exact test approach has a drawback in detecting significant residues since Fisher's test is very conservative especially when data are sparse.
For example, Fisher's test can handle some amino-acids which have strong evidence 
given that the residue frequencies in the two groups were extreme, however, likely important, other biological sites do not register under methods that depend on Fisher exact tests which becomes particularly limited when the differences in residue preference is either subtler or when the data are sparse.

To demonstrate this, we consider the studies in \cite{Deymier2015} which compare two groups of HIV-1 founder viruses based on their susceptibility to inactivation by vaginal microbiota acid metabolites. There are two groups and each group has a small sample size (Transmitted group  with 5 samples and Non-Transmitted group with 3 samples).
Each of sample consists of 900 sites, however some of them have missing values, so we consider 812 sites without missing values. Fisher's test with the procedure in \cite{BH} does not discover significant residue even under the nominal level of false discovery rate (FDR)=0.20 since the smallest p value is 0.0178 which is not considered as an significant site among 812 residues. Those residues, to a biologist’s eye, appear to differentiate the two phenotypes in the sense that the amino acids in most of the positions have chemical properties in the two groups that are different enough to possibly contribute to different functions, however none of sites is considered as an significant site under Fisher's test and Benjamini-Hochberg procedure. There are several reasons for such a gap between statistical result and biological interpretation which comes from inefficiency of Fisher's test and sparse count data due to small number of samples. 
 
The overall goal of this study is to propose methods for sparse count data, based on Bayesian approach which incorporate some scientific prior information into statistical models to reduce the gap between statistical significance and biological significance. In the first proposed model (section 4.1), we developed a Bayesian local false discovery with BLOSUM62 substitution matrix to identify significant residues distinguishing Transmitted and Non-Transmitted groups in HIV data. In the second proposed model (section 4.2), we proposed model for considering association among 20 amino acids. The proposed models leverage the BLOSUM62 substitute matrix in \cite{Robinson} and \cite{Heinkoff} as prior information on the likelihood of 20 amino acids while additionally considering information on dependencies among amino acids.

The development of new methods to detect significant sites in sequences have a broad impact on related areas. Outside of the HIV “donor vs. founder” and “founder vs. founder” types of comparisons, there are many other biological applications that could use an improved genetic signature scanning methodology. In general, any experimental setup susceptible to a two-independent samples test searching for genetic signatures or mapping phenotype to genotype can benefit from the improved sensitivity of the proposed approach.
For instance, vaccine and drug trials with their treated vs. placebo/untreated cohorts, studies analyzing disease-susceptible vs. disease-resistant groups, and any differential phenotype analyses.

This paper is organized as follows. In section 2, we present the motivation of our proposed models 
which are based on reflecting scientific knowledge through prior distributions in Bayesian model.  
Section 3 includes the overview of multiple testing procedures and 
the section 4 provide our two proposed methods such as empirical Bayes method and fully Bayes method with Markov random field model.  We discuss how we reflect the BLOSUM62 substitution matrix to our proposed models 
and provide the results from these two models and the comparison. 
We present conclusions and discussions in section 5. 

\section{Motivation of Proposed Method}
When data are sparse, it is desirable to develop an appropriate and efficient statistical testing procedures combining observed data and prior scientific knowledge to detect significant sites in sequence data. In detecting significant sites in HIV data, Fisher's exact test has limitation in detecting significant sites under sparse count data. 
There are several reasons that Fisher's exact test has such drawback: one reason is that the actual Type I error rate of the Fisher's exact test is far below the nominal level of Type I error. This is mainly from the discreteness of the Fisher's exact test and it is more serious for the conditional test such as Fisher's test than unconditional tests \cite{Agresti}. Another reason is that conditioning on the marginal total sums 
loses information which is demonstrated in \cite{Choi}. More serious problem is that Fisher's test considers only the configuration of counts, not different types of amino acids. This means that any multiple testing procedure with Fisher's exact test has the same decision on the same $p$-values regardless of types of amino acids. Some pairs of amino acids tend to occur more often than other pairs of amino acids, which may be more biologically relevant than commonly occurring pairs. Hence, it is more reasonable to consider rare pairs of amino acids more significantly than some common pairs of amino acids. Integrating such domain knowledge into statistical testing procedure could enhance the relevance of the analysis. 

We present an innovative strategy in the context of introducing biological knowledge from the BLOSUM62 substitute matrix and reflect the information on the decision of significance of sites through Bayesian models. One can judge such common and rare pairs of amino acids from the BLOSUM62 substitute matrix of which the scores provide information on how much often any pair of amino acids occurs. 
Therefore, it is more reasonable to take into account such information to make decision on significant sites. Based on information on association between phenotypes, \cite{Chung} considered Markov Random Field (MRF) to model the relationship among phenotypes on identifying SNPs with Genome-Wide Association Studies (GWAS) datasets. We consider the MRF model in the second method to generate a model that considers associations between amino acids.

The proposed study includes the direct application of the new methods to the same type of data sets which may be reported as non-discovery of significant sites from the traditional methods. As a result, the study not only addresses the limitations of existing approaches but also introduces an informed strategy that potentially advances our understanding of significant sites in biological contexts.  Overall, such biological knowledge or additional information will improve the performance of testing procedures  in detecting significant sites for sparse count data while traditional approaches such as Fisher's test has difficulty in finding out significant sites. Finally, this will be a first-time application to those data sets using the BLOSUM62 substitute matrix as a prior information.


\section{Overview of Multiple Testing procedures controlling FDR}
In this section, we present a brief overview of multiple testing procedures to control the false discovery rate (FDR). 
In hypothesis testing, a common problem arises where numerous hypotheses are evaluated simultaneously. This leads to the multiple testing problem. Consider we test $n$ 
null and alternative hypotheses such as 
\begin{eqnarray}
 H_{0i} :  \quad vs. \quad H_{1i}
 \label{eqn:hypotheses}
\end{eqnarray}
for $i=1,\mathellipsis,n$ at the same time. We need careful attention to conclude which one to reject from the $p$-values of the individual tests. To address this challenge, several approaches such as family-wise error rate (FWER) control and false discovery rate (FDR) control have been developed. 

First, the probability of rejecting at least one null hypothesis when all null hypotheses are true is known as FWER which is defined as  
\begin{align*}
    FWER = P\{\text{reject any true null hypothesis } H_{0i}\}.
\end{align*} 
Since the FWER methods control the probability that the number of false positives is greater than zero, the result is quite conservative 
in large scale multiple testing problems such as 
thousands or tens of thousands of hypotheses.  Therefore, the FWER is not used in  
such large scale multiple testing problems due to the lack of testing powers. 
Rather than using the FWER,  another error rate in multiple testing is the false discovery rate (FDR) 
which is quite popular in large scale problems. 
More specifically, let $\mathcal{D}$ be a decision rule and define $V$, 
\begin{eqnarray}
V=\sum_{i=1}^n I(\mbox{$H_{0i}$ is rejected, but $H_{0i}$ is true})
\end{eqnarray}
as the number of falsely rejected null hypotheses and 
$S = \sum_{i=1}^n I(\mbox{$H_{0i}$ is rejected and $H_{1i}$ is true})$ as the number of correctly rejected null hypotheses. So the $R = V+S$ null hypotheses out of $n$ null hypotheses are  rejected. 
Then the false discovery proportion (FDP) is defined as 
\begin{align*}
    FDP (\mathcal{D}) = \frac{V}{\text{max}(R, 1)}
\end{align*}
and the FDR is 
\begin{eqnarray}
FDR(\mathcal{D})=E(FDP (\mathcal{D})). 
\end{eqnarray}

\cite{BH} proposed 
a procedure which controls the FDR for independent $p$-values as follows.  
For the $n$  independent null hypotheses and 
the corresponding $p$-values denoted by  $p_1,...,p_n$,  the ordered $p$-values are  $p_{(1)} \leq \dots \leq p_{(n)}$. 
For a given level of FDR $\alpha \in (0,1)$,
define $i_{\max} = \max\left\{i: p_{(i)} \leq \frac{i}{n}\alpha \right\}.$ Then the decision rule $\mathcal{D}_\alpha = \{\text{Reject all} ~~ H_{(i)} ~~\text{for $i$ } \leq i_{\max}\} $ has $FDR(\mathcal{D}_\alpha)\leq \alpha$.
While \cite{BH} validated this procedure for independent $p$-values, 
\cite{BY} generalized this procedure to handle the case with dependent $p$-values such as positive regression dependent. In addition, the procedure in \cite{BH} is also valid for discrete test statistics as well for appropriately defined $p$-values. In our context,  the $p$-values computed from Fisher's exact test with the procedure in \cite{BH} is expected to control a given level of FDR.    

In addition to the procedure in \cite{BH} based on $p$-values, 
a simple Bayesian approach for multiple testing is also suggested by \cite{Efron} which  introduced local false discovery rate (lfdr). 
Each of the $n$ cases belongs to either null or non-null with a certain prior probability. Let the prior probabilities be $\pi_0$ and $\pi_1=1-\pi_0$, respectively. (i.e. $\pi_0 = \frac{n_0}{n}$ is the proportion of true null where $n_0$ is the number of the true $H_{0i}$.) Corresponding prior densities  for the observation or test statistic $x$ are  $f_0(x)$ and $f_1(x)$, respectively. 
Then the resulting mixture density, denoted as $f(x)$, can be expressed as $f(x) = \pi_0f_0(x)+\pi_1f_1(x)$. Applying Bayes' theorem, the posterior probabilities are given as follows:
\begin{align*}
    &Pr(\text{$H_{0i}$ is true }|X_i=x) = \frac{\pi_0f_0(x)}{f(x)}\\
    &Pr(\text{$H_{1i}$ is true }|X_i=x) = 1-\frac{\pi_0f_0(x)}{f(x)}    
\end{align*}

Then, the Bayesian FDR for ${X \leq x}$,  also called $q$ value in \cite{Storey}, is $FDR(x) =Pr\{\text{$H_{0i}$ is true }| X_i \leq x\} = \frac{\pi_0F_0(x)}{F(x)}$, where $F_0(x)$ and $F(x)$ be the cumulative distribution function values corresponding to $f_0(x)$ and $f(x)$. While the $q$-value is the tail-area false discovery rate, the local false discovery rate ($lfd r$) consider the probability of null given $X_i=x_0$ as follows: 
$$lfdr(x_0) = Pr\{\text{$H_{0i}$ is true }| X_i = x_0\}=\frac{\pi_0 f_0(x_i)}{f(x_i)}.$$
In order to compute the lfdr, 
we need to compute three components :
the null proportion ($\pi_0$), 
the null density ($f_0$) 
and the marginal density  ($f = \pi_0f _0 + (1-\pi_0) f_1$). 

For HIV data, we use  
 parametric probability functions such as multinomial distributions  
 and their conjugate prior  distributions for parameters, Dirichlet distributions, 
 our statistical inference is  based on sampling from Monte Carlo Markov Chain (MCMC).  
For some parameters such as hyperparameters in prior distributions, 
one of our proposed methods adopts
the Empirical Bayes (EB) in \cite{Morris} to estimate those hyperparameters by maximizing the marginal likelihood of ${\bf x}$, 
$\hat \alpha = \arg \max_{\alpha} m({\bf x}|\alpha)  $ 
where 
$m({\bf x}) = \int f({\bf x}|\theta) \pi(\theta|\alpha) d\theta$ 
for the likelihood $f({\bf x}|\theta)$ and the prior distribution $\pi(\theta|\alpha)$. 

\section{Proposed Methods}
In this section, we propose two methods to analyze 
the HIV data considered in \cite{Deymier2015}.  
One method is based on independent   
assumptions of  amino acids using the EB method  
and the other is fully Bayesian approach  
using Markov Random Field (MRF) model with 
pairwise dependence of amino acids.
The first model is simpler than the second model, so 
the first one can be implemented more easily while 
the second one provides more delicate model at the cost of heavy computations. 
Both models reflect the BLOSUM62 substitute matrix, however 
the first one uses only the diagonal terms while the second model 
utilize off-diagonal terms which carry out the information on pairwise association of 20 amino acids. 

We first present our baseline model for the distributions of the counts of  amino acids in each site.  Suppose there are two groups (Transmitted (T) and Non-Transmitted (NT)) and each group has $n_1$ and $n_2$ sample vectors where each sample vector consists of $K$ sites. Each site $i$ for $1\leq i \leq K$ has multinomial distributions for Transmitted and Non-Transmitted group which are  
\begin{eqnarray}
\textbf{x}_i &\sim& \rm Multinomial(n_1, \textbf{p}_{T}), \\
\textbf{y}_i &\sim& \rm Multinomial(n_2, \textbf{p}_{NT})
\end{eqnarray}
 where $\textbf{p}_T$ and $\textbf{p}_{NT}$ are 20 dimensional probability vectors for $20$ kinds of amino acids for each groups. The goal is to test the following $K$ multiple hypotheses for $K$ sites:
we test 
\begin{eqnarray*}
H_{0i} &:& \mbox{distribution of amino acids in $T$ and $NT$ groups are the same in $i$th site}\\
H_{1i} &:& \mbox{ not $H_{0i}$ }
\end{eqnarray*}
for $1\leq i \leq K$
by controlling some type I error rate such as false discovery rate (FDR). 
The small $p$-values from the Fisher's exact test are  
presented in  Table \ref{tbl:preliminary}. The BH procedure with those $p$-values from Fisher's exact test does not reject any of them.   

We use the idea of Bayesian local false discovery rate which is also known as $q$-value based on mixture of two distributions. For each site $i$, ${\bf x}_i$ and ${\bf y}_i$ have the common probability vector
${\bf p}_i \equiv {\bf p}_{T} = {\bf p}_{NT}$ under the null hypothesis while
${\bf p}_T$ and ${\bf p}_{NT}$ are different under the alternative hypothesis.
Let ${\bf z}_i =({\bf x}_i, {\bf y}_i)$, then we have the marginal distribution of ${\bf z}_i$;
\begin{eqnarray*}
f({\bf z}_i) &=& \pi_0 f_0({\bf z}_i) +(1-\pi_0) f_1({\bf z}_i)
\end{eqnarray*}
where $\pi_0$ is the mixing proportion of the null hypotheses, and
\begin{eqnarray*}
f_0({\bf z}_i) &=&  \int h_{T}({\bf x}_i|{\bf p}) h_{NT}({\bf y}_i|{\bf p}) dG_{0}({\bf p}), \\
f_1({\bf z}_i) &=&  \int \int h_T({\bf x}_i|{\bf p}_T) dG_T({\bf p}_{T}) h_{NT}({\bf y}_i| {\bf p}_N) dG_N({\bf p}_N),
\end{eqnarray*}
where $h_T({\bf x}|{\bf p}_{T}) = \binom{n_1}{ x_1, x_2,\ldots, x_s}\prod_{i=1}^{s} p_{T,i}^{x_i}$ which is the multinomial distribution
for ${\bf p}_{T}=(p_{T,1},\ldots, p_{T,20})$ and $h_{NT}({\bf y}|{\bf p}_{NT})$ is also similarly defined. The prior distribution of $\textbf{p}_{T}$ and $\textbf{p}_{NT}$ are defined as
\begin{align}
    & \textbf{p} \equiv \textbf{p}_{T} \equiv \textbf{p}_{NT} \sim \rm Dirichlet(\alpha_1,...,\alpha_{20}) \text{ under $H_{0i}$}, \label{eqn:Diri_null}  \\
    & \textbf{p}_{T} \sim \rm Dirichlet(\alpha_1,...,\alpha_{20}), \textbf{p}_{NT} \sim \rm Dirichlet(\alpha_1,...,\alpha_{20}) \text{ under $H_{1i}$}
    \label{eqn:Diri_alt}
\end{align}
which are the conjugate priors for multinomial distributions.

\subsection{Model 1 : Empirical Bayes }
The choice of the hyperparameter $\boldsymbol{\alpha} = (\alpha_1,\mathellipsis, \alpha_{20})$ in \eqref{eqn:Diri_null} and \eqref{eqn:Diri_alt}  is a crucial issue in this analysis since the results may be  sensitive to the choice of  $\boldsymbol{\alpha}$. 
For Dirichlet distribution, there are several options, for example, Jeffrey's prior in \cite{Jeffreys} and reference prior in \cite{Berger} which use $\alpha_i=1/2$ and $\alpha_i=1/K$, respectively. Under Jeffrey's prior and reference prior, we do not reject any site when the nominal level FDR is  0.05. 
Instead of these priors, we proposed to 
borrow the information on the probabilities for 20 amino acids from the BLOSUM62 substitution matrix \cite{Robinson} and  \cite{Heinkoff}. BLOSUM (BLocks SUbstitution Matrix) matrix is a substitution matrix used to score the similarity between amino acids in protein sequences. 
From the BLOSUM62 substitution matrix, the probabilities of 20 amino acids, denoted by ${\bf q}=(q_1,\ldots, q_{20})$, can be derived and it can be used in the construction of prior distributions as follows. 
From Dirichlet distribution in \eqref{eqn:Diri_null} and \eqref{eqn:Diri_alt}, since  $E(p_s) = \frac{\alpha_s}{\sum_{s=1}^{20} \alpha_s}$  from $\rm Dir(\alpha_1,\ldots, \alpha_{20})$, we match the moments of $p_s$ with  the probability vector ${\bf q}=(q_1, \ldots, q_{20})$ obtained from BLOSUM62.  
{The derivation of ${\bf q}$ is presented in Appendix \ref{appendixA}.}  This motivates to use $\alpha_s =  \beta q_s$ for some $\beta>0$ since $E(p_s) \propto \alpha_s$, therefore 
we now have one hyper-parameter $\beta>0$  instead of 20 $\alpha_s$s. 
Through the  utilization of the BLOSUM62 substitution matrix, our approach integrates  biological information into our model, thereby diminishing the number  of  parameters to be estimated significantly. 
This not only enhances the efficiency of our model but also emphasizes our commitment to incorporating  scientific knowledge to statistical model which leads to  more scientifically valid results.

Since the results are sensitive to the choice of $\beta$ we use a more objective way such as  the idea of empirical Bayes (EB) approach to maximize
the marginal likelihood of the data \cite{Morris}. 
There are 3 values of $\beta$: one is for the null distribution ($\beta_0$) 
in \eqref{eqn:Diri_null} and the other two
are for alternative hypothesis in \eqref{eqn:Diri_alt}  which are denoted by  $\beta_T$s and $\beta_{NT}$ for ${T}$ and ${NT}$ groups, respectively. 
The EB estimators are obtained from maximizing the marginal likelihoods 
of ${\bf x}_{i}$ and ${\bf y}_{i}$. 
Depending on the null and alternative hypotheses, we have different forms of marginal likelihood functions. 

Under the null hypothesis, 
the marginal likelihoods of $\textbf{z}_i 
=(\textbf{x}_i, \textbf{y}_i)$ 
with $\textbf{x}_i=(x_{ij})_{1\leq j \leq 20}$ and $\textbf{y}_i=(y_{ij})_{1\leq j \leq 20}$, $P(\textbf{z}_i|\boldsymbol{\alpha})$ (Dirichlet-Multinomial probability mass function for $\textbf{z}_i$) is defined as follows:
\begin{align*}
  & \int_{\bf p} h_0(\textbf{z}_i|{\bf p})G_0({\bf p}|\boldsymbol{\alpha})d{\bf p}
  \\ & = \int \prod_{j=1}^{20}p_{ij}^{x_{ij}}p_{ij}^{y_{ij}}\frac{\Gamma(\sum_{j=1}^{20}\alpha_j)}{\prod_{j=1}^{20}\Gamma(\alpha_j)}\prod_{j=1}^{20}p_{ij}^{\alpha_j-1}d{\bf p} =\frac{\Gamma(\sum_{j=1}^{20}\alpha_j)\prod_{j=1}^{20}\Gamma (\alpha_j+x_{ij}+y_{ij})}{\prod_{j=1}^{20}\Gamma (\alpha_j)\Gamma(n_1+n_2+\sum_{j=1}^{20}\alpha_j)}
\end{align*}

Under the alternative hypothesis, $\textbf{p}_{T}$ and $\textbf{p}_{NT}$ are different and two groups are considered to follow Dirichlet distribution independently, so the likelihoods is defined as follows:
\begin{align*}
     & \int_{\bf p_T} h_T({\bf x_i}|{\bf p_T})G_T({\bf p_T}|\boldsymbol{\alpha})d{\bf p_T} \times \int_{\bf p_N} h_{NT}({\bf y_i}|{\bf p_N})G_N({\bf p_N}|\boldsymbol{\alpha})d{\bf p_N} 
     \\ & = \frac{\Gamma(\sum_{j=1}^{20}\alpha_j)\prod_{j=1}^{20}\Gamma (\alpha_j+x_{ij})}{\prod_{j=1}^{20}\Gamma (\alpha_j)\Gamma(n_1+\sum_{j=1}^{20}\alpha_j)}\times \frac{\Gamma(\sum_{j=1}^{20}\alpha_j)\prod_{j=1}^{20}\Gamma (\alpha_j+y_{ij})}{\prod_{j=1}^{20}\Gamma (\alpha_j)\Gamma(n_2+\sum_{j=1}^{20}\alpha_j)}
\end{align*}

As previously established, we introduced the parametrization of $\alpha_j=\beta q_j$, where $\sum_{j=1}^{20}q_j=1$ by employing the frequency-based BLOSUM substitution matrix. Then the EB estimators are obtained from maximizing the marginal likelihood of ${\bf z}_i$ which is

\begin{align*}
\hat \beta_0 = {\rm argmax}_{\beta>0}  \prod_{i=1}^{K} h_0({\bf z_i}|\boldsymbol{\alpha})
= {\rm argmax}_{\beta>0}  \prod_{i=1}^{K} \frac{\Gamma(\beta) \prod_{j=1}^{20}\Gamma(\beta q_j + x_{ij}+ y_{ij} )}{ \prod_{j=1}^{20} \Gamma(\beta q_j) \Gamma(n_1 + n_2 +\beta)}.
\end{align*}
Similarly,
\begin{align*}
\hat \beta_T
= {\rm argmax}_{\beta>0}  \prod_{i=1}^{K} \frac{\Gamma(\beta) \prod_{j=1}^{20}\Gamma(\beta q_j + x_{ij})   }{ \prod_{j=1}^{20} \Gamma(\beta q_j) \Gamma(n_1 +\beta)},~~~
\hat \beta_N
= {\rm argmax}_{\beta>0}  \prod_{i=1}^{K} \frac{\Gamma(\beta) \prod_{j=1}^{20}\Gamma(\beta q_j + y_{ij})   }{ \prod_{j=1}^{20} \Gamma(\beta q_j) \Gamma(n_2 +\beta)  }.
\end{align*}

The local false discovery rate (lfdr) at ${\bf z}_i$ is $ lfdr({\bf z}_i) = \frac{\pi_0 f_0(z)}{\pi_0 f_0(z) +(1-\pi_0) f_1(z)}$ and the hypothesis $H_{0i}$  is rejected if $lfdr({\bf z}_i) \leq \alpha$ for  any given nominal level of FDR $\alpha$, typically $\alpha=0.05$. To estimate the local false discovery rate, the estimation of the null proportion is needed. We introduced a latent indicator variable $e_i$ where it takes a value of 1 if group T and NT are different in i-th site and 0 otherwise, and $Unif(0,1)$ as our prior for $\pi_0$. Then we used Metropolis Hastings within Gibbs sampling framework with the following conditional distributions:

\begin{algorithm}
    \caption{Metropolis Hastings within Gibbs Sampling for the null proportion}
    \begin{algorithmic}[1]
        \State Choose initial value for $\pi_0$ and $\mathbf{e}$
        \For{$m=1$ to $M$}
            \State Sample $\pi_0^{(m)}$ $\sim$ $f(\pi_0|\mathbf{e}^{(m-1)}, \mathbf{x, y, p})$ $\sim \text{Beta}(n - \sum_{i}e_i + 1, \sum_{i}e_i + 1)$
            \State Sample $\mathbf{e}^{(m)}$ $\sim$ $f(e|\pi_0^{(m)})$ $\sim \text{Bernoulli}(1- \text{lfdr}(z_i))$ $\text{, where } \text{lfdr}(z_i)=\frac{\pi_0^{(m)} f_0(z_i)}{f(z_i)}$
        \EndFor
    \end{algorithmic}
\end{algorithm}

Through Gibbs sampling based on the conditional probability, we processed sampling M times (Here, we used M = 10000.). Then $lfdr ({\bf z}_i)$ is estimated by averaging lfdr from M sampling as $ lfdr({\bf z}_i) \approx \frac{1}{M} \sum_{m=1}^M  lfdr( {\bf z}_i |  \Theta_m)$ where $\Theta_m = (  {\bf p}_T, {\bf p}_{NT}, {\bf p})$ are generated from the posterior distributions at $m$th sampling. 

We obtained  $\hat \beta_0=0.2596$, $\hat \beta_T=0.2730$ and $\hat \beta_N=0.0648$. Using these values, we reject 26 sites as in Table \ref{tbl:preliminary}.
\begin{table}[hbt!]
\begin{center}
\begin{tabular}{cccccc}
position& Fisher's & local-fdr&	  	Transmitted&	Non- & Score \\
        & $p$-value  & $q$-value  &                &     transmitted &                \\
\hline
142	 &0.0892    &  0.0429	&D1 K1 F1 P2	&H3      & DH=-1, KH=-1, FH=-1, PH=-2   \\
759	 &0.0892    &  0.0429	&N1 D1 Q2 T1	&H3       & NH=1, DH=-1, QH=0, TH==-2 \\
7	 &0.0535    &  0.0451	&D2 K1 P2	&Q3       & DQ=0, KQ=1, PQ=-1 \\
262	 &0.0535    &  0.0454	&E2 P1 V2	&I3       & EI=-3, PI=-3, VI=3 \\
585	 &0.0535    &  0.0455 	&G1 P2 V2 &I3      &  GI=-4, PI=-3, VI=3  \\
245	 &0.0178    &  0.0472 	&K1 T4 &H3      & KH=-1, TH=-2  \\
627	 &0.0357    &  0.0473 	&K3 S2 &T3      & KT=-1, ST=5  \\
327	 &0.0178    &  0.0473 	&K4 S1 &T3      & KT=-1, ST=5  \\
222	 &0.0357    &  0.0473	&H3 P2 &F3       & HF=-1, PF=-4 \\
111	 &0.0178    &  0.0473	&G1 T4 &S3       & GS=0, TS=1 \\
138	 &0.0178    &  0.0474	&N1 K4	&E3       & NE=0, KE=1 \\
776	 &0.0178    &  0.0474	&N4 T1	 &D3    & ND=1, TD=-1  \\
449	 &0.0357    &  0.0474	&A3 T2	 &W3    & AW=-3, TW=-2\\
749	 &0.0178    &  0.0476	&A4 W1	 &G3    & AG=0, WG=-2 \\
699	 &0.0178    &  0.0477	&G4 W1	 &I3    & GI=-4, WI=-3 \\
738	 &0.0178    &  0.0478	&H4 P1	 &L3    & HL=3, PL=-3 \\
463	 &0.0178    &  0.0493	&I5	 &M3    & IM=1\\
879	 &0.0178    &  0.0495	&P5	 &H3    & PH=-2\\
492	 &0.0178    &  0.0496	&A5	 &S3    & AS=1 \\
894	 &0.0178    &  0.0496	&D5	 &F3    & DF=-3 \\
582	 &0.0178    &  0.0496	&K5	 &S3    & KS=0 \\
617	 &0.0178    &  0.0498 	&N5   &D3  & ND=1 \\
235	 &0.0178    &  0.0498	&A5	&W3    & AW=-3 \\
389	 &0.0178    &  0.0499	&G5	&W3    & GW=-2 \\
491	 &0.0178    &  0.0499	&I5	&G3    & IG=-4 \\
893	 &0.0178    &  0.0499	&I5	&G3    & IG=-4 \\
\hline
\end{tabular}
\caption{\label{tbl:preliminary} Preliminary results. 26 significant sites rejected from local fdr ($q$-value) and $p$-values from Fisher's test.
  }
\end{center}
\end{table}
Rejection sites include 10 of $\{(5,0),(0,3)\}$, 8 of $\{(4,1,0),(0,0,3)\}$, 3 of $\{(3,2,0),(0,0,3)\}$, 3 of $\{(2,2,1,0),(0,0,0,3)\}$ and 2 of $\{(2,1,1,1,0),(0,0,0,0,3)\}$, while Fisher's test with Benjamini-Hochberg procedure does not reject any site when $\alpha=0.05$. Table \ref{tbl:preliminary} shows the $p$-values from Fisher's test and local fdr values for 26 significant sites. 


\subsection{Model 2 :  Considering Dependency Among Pairwise Amino Acids and MRF Model}
In the previous section on EB approach, we consider the information on  
each of amino acids  in the sense that 
we construct prior distribution of 20 amino acids through  using 
only the probability vector $\textbf{q}$ corresponding to 
20 amino acids  derived from the diagonal terms in BLOSUM62 substitution matrix.
This approach tends to have almost the same decision on the sites with the same configuration of counts. However, Table \ref{tbl:preliminary} shows that the scores corresponding to amino acid pairs are different across different residues although they are considered all significant from the EB approach. 
Although the same configurations of the counts are observed, different amino acid types with the same counts may have different meanings. In other words, an occurrence of a pair of rare amino acids should be considered more importantly than common pairs of amino acids. The prior information on rare amino acids can be obtained from the BLOSUM62 substitute matrix which consists of logarithms of odds of two amino acids. The scores in off-diagonal in the BLOSUM62 matrix range from -4 to 2 and the majority of scores are negative values. Those scores represent that each pair of amino acids with a smaller score occurs less often than those with higher scores. In the EB approach in the previous section, significant residues are selected without considering information on the likelihood of pairwise amino acids. If the likelihood of pairwise amino acids is important in detecting significant sites, it is meaningful to incorporate such information from the BLOSUM62 substitute matrix on those likelihoods of pairwise amino acids into statistical models. 

\subsubsection{Proposed Method}
To model the dependency among pairwise amino acids considering the BLOSUM62 substitution matrix, we used the idea of Markov Random Field (MRF) (\cite{Besag}, \cite{Wei}). 
For the observational vectors ${\bf x}_i$ and ${\bf y}_i$ in $i$th residue from Transmitted and Non-Transmitted groups, let ${\bf z}_i=({\bf x}_i,{\bf y}_i$). 
Under the null hypothesis, the probability distribution function of ${\bf z}_i$ $(i=1,...,K)$ is modeled as
\begin{align}
    P({\bf z}_i|\Theta_s, \Theta_{st}, \delta)=\exp\left(\sum_{s=1}^{20}\theta_s x_{is}+\delta \sum_{t \neq s}^{}\theta_{st}z_{is}z_{it} \right)\cdot C(\Theta), \label{eqn:MRF}
\end{align}
where $ C(\Theta)$ is a normalizing constant, $\delta$ is a  parameter 
which determines the magnitude of effect of pairwise amino acids, and 
\begin{eqnarray}
\Theta_s &=& (\theta_1,...,\theta_{20}), \label{eqn:null_Theta_s}\\
\Theta_{st}&=&(\theta_{1,2}, ...,\theta_{19,20})
\label{eqn:null_Theta_st}
\end{eqnarray}
 are coefficients for independent and pairwise terms, respectively.  For alternative hypothesis, we similarly define $P({\bf z}_i|\Theta_s^{T}, \Theta_{st}^{T}, \delta)$ and $P({\bf y}_i|\Theta_s^{N}, \Theta_{st}^{N}, \delta)$
 where  the parameter vectors $\Theta_s^T, \Theta_{st}^T$ and $\Theta_s^N, \Theta_{st}^N$ which are also similarly defined as in \eqref{eqn:null_Theta_s} and \eqref{eqn:null_Theta_st}.

Since the normalizing constant $C(\Theta)$ makes computationally difficult to compute the full likelihood function, one common approach is the pseudo-likelihood\cite{Besag} which is
\begin{align*}
    \text{Pseudo-likelihood of } {\bf z}_i = & \prod_{s=1}^{20} P(z_{is}|{\bf z}_{i,-s}, \Theta_s, \Theta_{st}, \delta)
    \end{align*}
    where 
    \begin{align*}
    & P(z_{is}|{\bf z}_{i,-s},\Theta_s, \Theta_{st}, \delta) = \exp\left(\theta_sz_{is}+\delta \sum_{t \neq s}^{}\theta_{st}z_{is}z_{it} \right)\cdot C(\Theta_{-s}),\\
    & C(\Theta_{-s}) =\sum_{z_{is} \in \{0,1,...,8\}} \exp\left(\theta_s z_{is}+\delta \sum_{t \neq s}^{}\theta_{st}z_{is}z_{it}\right),
\end{align*}
 ${\bf z}_i=(z_{i1},...,z_{i20})$ and ${\bf z}_{i,-s}=(z_{i1}...z_{i,s-1}, z_{i,s+1},...,z_{i,20})$ is the vector excluding $z_{is}$. The pseudo-likelihood of ${\bf x}_1,...,{\bf x}_K$ and ${\bf y}_1,..,{\bf y}_K$ for model under non-null is similarly defined with the parameter vectors $\Theta_s^T, \Theta_{st}^T$ and $\Theta_s^N, \Theta_{st}^N$ which are also similarly defined as in \eqref{eqn:null_Theta_s} and \eqref{eqn:null_Theta_st}.

Under this model, we define $\theta_s=\log(p_s)$ and 20 dimensional vector $\mathbf{P}_s$
\begin{eqnarray}
\mathbf{P}_s=(p_1,\ldots, p_{20}),
\end{eqnarray}
and define  $\theta_{st}=\log(p_{st})$ and $r=\frac{20(20-1)}{2}$ dimensional vector  $\mathbf{P}_{st}$ which 
is a vectorization of $p_{st}$ for $1\leq s< t \leq 20$  
\begin{eqnarray}
\mathbf{P}_{st}=(p_{12},p_{13},\ldots, p_{19,20} ).
\end{eqnarray}
Similar to the model 1, $p_s$ is positive probability value with the constraint $\sum_{s=1}^{20}p_s=1.$ 
{Additionally, $p_{st}$ is also positive probability value with the constraint $2\sum_{1 \leq s<t \leq 20}p_{st}=1$ where we further assume that coefficients for independent and pairwise therm are not dependent to each other.}

For Bayesian inference, we give Dirichlet distribution as a prior distribution to $\mathbf{P}_{s}$ and $\mathbf{P}_{st}$, 
\begin{eqnarray}
\mathbf{P}_{s} &\sim& Dirichlet(\alpha_1,...,\alpha_{20}), \\
\mathbf{P}_{st} &\sim& Dirichlet(\alpha'_1,...,\alpha'_r).
\end{eqnarray}
As in the case of the EB approach, we have $E(p_s)=\frac{\alpha_s}{\sum_{s=1}^{20}\alpha_s}=q_s$ and $E(p_{st})=\frac{\alpha'_l}{\sum_{l=1}^{r}\alpha'_l}=q_{st}$, where $q_s$ is the probability of each amino acids (Recall the probabilities of 20 amino acids, ${\bf q}$ in Model 1) and $q_{st}$ is the pairwise probability obtained from the BLOSUM62 substitution matrix.  
{The derivation of $q_s$ and $q_{st}$ are attached in Appendix \ref{appendixA}.} 
This motivates to take $\alpha_s=\beta_1 q_s$ and $\alpha'_l=\beta_2 q_{st}$ for some $\beta_1$ and $\beta_2.$
We predetermined these hyperparameters as $\beta_1=1000$ and $\beta_2=10000$, and we will provide discussion about this setting in section 5. 

In the model, there are parameters $\mathbf{P}_s, \mathbf{P}_{st}$ need to be estimated. We generate samples of ($\mathbf{P}_s, \mathbf{P}_{st}$) from posterior distributions of $p_s$ and $p_{st}$ using MCMC (Monte Carlo Markov Chain). We used the characteristic that $p_s$ has a Beta distribution with parameters $\alpha=\alpha_s$ and $\beta=\alpha_1+\ldots+\alpha_{s-1}+\alpha_{s+1}+\ldots+\alpha_{20}$, so that $p_1, ...,p_{20}$ have a Dirichlet distribution with parameters $\alpha_1,...,\alpha_{20}$. Similarly, for $p_{st}$, let $p_{st} \sim Beta(\alpha, 
\beta)$ where $\alpha=\alpha'_{st}$ and $\beta=\sum_{l=1}^{r}\alpha'_l-\alpha'_{st}$, so that $p_{1,2}, ...,p_{19,20}$ have a Dirichlet distribution with parameters $\alpha'_1,...,\alpha'_r$.
Prior to sampling, we compared the target density for each $\frac{20(20+1)}{2}$ parameters with normal distribution where parameters were obtained by moment matching by Laplace approximation\cite{Laplace}. We set the initial values closer to actual target parameters through this step in order to proceed efficient sampling. Algorithm 2 shows the MCMC algorithm for the posterior inference.

\begin{algorithm}
    \caption{MCMC algorithm for posterior inference of Model 2 Parameters }
    \begin{algorithmic}[1]
    
        \State Obtain initial values of $\mathbf{P}_s, \mathbf{P}_{st}$ from BLOSUM62 substituion matrix. 
        
        \For{$s=1$ to $20$} 
        \State Propose $p_s^{*}$ from proposal density $q(p_{s}^{*}|p_{s}^{(m-1)}) \sim N(p_{s}^{(m-1)}, \sigma^2)$.
        \State Calculate the ratio R as follows:
        \begin{align*}
        R &=\frac{f(p_s^*| \mathbf{P}_{-s}^{(m-1)})}{f(p_s^{(m-1)}| \mathbf{P}_{-s}^{(m-1)})}\\
        &=\prod_{i=1}^{K} \frac{C(p_s^*,  \mathbf{P}_{-s}^{(m-1)}) \cdot{\rm  exp(log}(p_s^{*})z_{is}))\times {\rm Beta}(\alpha_s^*, \sum_{j \neq s}\alpha_j^{(m-1)} )}{C(p_s^{(m-1)},  \mathbf{P}_{-s}^{(m-1)}) \cdot {\rm exp(log}(p_s^{(m-1)})z_{is}))\times {\rm Beta}(\alpha_s^{(m-1)}, \sum_{j \neq s}\alpha_j^{(m-1)})},\\
        & \text{where $\alpha_s^{(m-1)}=\beta_1 p_s^{(m-1)}$ and $\alpha_s^*= \beta_1 p_s^{*}$.}
        \end{align*}
        \State Update $p_s^{(m)}=p_{s}^{*}$ with acceptance probability $min\{1, R\}$.
        \State Scale $(p_1^{(m)},...,p_{20}^{(m)})$ so that the sum of the mth block becomes equal to 1.\\
        \EndFor
        \\
        \For{$s=1,...,20, 1 \leq s < t \leq 20$}
        \State Propose $p_{st}^{*}$ from proposal density $q(p_{st}^{*}|p_{st}^{(m-1)}) \sim N(p_{st}^{(m-1)}, \phi^2)$.\\
        \State Calculate the ratio R as follows:
        \begin{align*}
        R &=\frac{f(p_{st}^*|\mathbf{P}_{-st}^{(m-1)})}{f(p_{st}^{(m-1)}|\mathbf{P}_{-st}^{(m-1)})}\\
        &=\prod_{i=1}^{K} \frac{C(p_{st}^*, \mathbf{P}_{-st}^{(m-1)}) \cdot {\rm exp}(\delta \cdot {\rm log}(p_{st}^{*})z_{is}z_{it}))\times {\rm Beta}(\alpha_{st}^*, \sum_{j \neq \{st\}}\alpha_j^{(m-1)} )}{C(p_{st}^{(m-1)},\mathbf{P}_{-st}^{(m-1)}) \cdot {\rm exp}(\delta \cdot {\rm log}(\theta_{st}^{(m-1)})z_{is}z_{it}))\times {\rm Beta}(\alpha_{st}^{(m-1)}, \sum_{j \neq \{st\}}\alpha_j^{(m-1)})}, \\
        & \text{where $\alpha_{st}^{(m-1)}=\beta_2 p_{st}^{(m-1)}$ and $\alpha_{st}^*=\beta_2 p_{st}^{*}$.}
        \end{align*}
        \State Update $p_{st}^{(m)}=p_{st}^{*}$ with acceptance probability $min\{1, R\}$.
        \State Scale $(p_{1,2}^{(m)},...,p_{19,20}^{(m)})$ so that the sum of the mth block becomes equal to 1.\\
        \EndFor
    \end{algorithmic}
\end{algorithm}

We fixed $\sigma^2$ with the median of initially obtained variance of $(p_1,..., p_{20})$ and $\phi^2$ with the median of initially obtained variance of $(p_{1,2},..., p_{19,20})$. In the Algorithm2, the reason for scaling is based on the assumption that $(p_1,...,p_{20}$) follows a Dirichlet distribution.
We generated samples from 10,000 MCMC iterations. Then we compute the local false discovery rate of ${\bf z}_i=({\bf x}_i, {\bf y}_i)$ with samplings of parameters at mth iteration in MCMC procedure $\mathbf{P}^m, \mathbf{P}_T^m, \mathbf{P}_N^m$, which is 
\begin{align*}
    & lfdr({\bf z}_i) =  \frac{\pi_0f_0({\bf x}_i, {\bf y}_i)}{\pi_0f_0({\bf x}_i, {\bf y}_i)+\pi_1 f_1({\bf x}_i)f_1({\bf y}_i)} \approx  \frac{\pi_0\widehat{f_0}({\bf x}_i, {\bf y}_i)}{\pi_0\widehat{f_0}({\bf x}_i, {\bf y}_i)+\pi_1 \widehat{f_1}({\bf x}_i)\widehat{f_1}({\bf y}_i)},
\end{align*}
where $\pi_0\widehat{f_0}({\bf x}_i, {\bf y}_i)\approx \frac{1}{M}\sum_{m=1}^{M}\widehat{\pi_0}f_0({\bf z}_i|\mathbf{P}^m)$, $\pi_1\widehat{f_1}({\bf x}_i)\widehat{f_1}({\bf y}_i)\approx \frac{1}{M}\sum_{m=1}^{M}\widehat{\pi_1} f_1({\bf z}_i|\mathbf{P}_T^m, \mathbf{P}_N^m)$, and $\pi_0 \approx \widehat{\pi_0}$ obtained from EB model. The ith site is considered to be a significant site if $lfdr({\bf z}_i) \leq \alpha$ when $\alpha$ is a given level of false discovery rate. 

\subsubsection{Result}
We select $\delta=1$, $\delta=0.5$ and $\delta=0.1$ and compare the effect of the pairwise terms. With FDR control level $\alpha=0.05$, we rejected 79 sites when $\delta=0.1$ and 112 sites when $\delta=0.5$. Note that the value of $\delta$ can be viewed as a weight for the effects of pairwise amino acids, and as the $\delta$ increases, Model 2 tends to more accurately reflect interactions, leading to a tendency to select  a larger number of sites.
In particular, we confirmed that 26 sites identified by Model 1 were also identified by Model 2 considering pairwise terms, and were located in the top rank when the sites were sorted according to the $p$-values in increasing order. 

In order to demonstrate the effect of BLOSUM62 score in detection when using Model 2, we obtained independent and pairwise scores for each site and evaluate the pattern between scores and detection sites. Since there are distinct constructions of each site in the data, an appropriate measure for summarizing score is required to compare each site based on their scores. The number of occurrences of counts among 20 amino acid positions for each site can range from 1 to 8. Based on this fact, it seems reasonable to put weight on independent scores over pairwise scores if the number of observed amino acid types is small, and weight on pairwise scores over independent scores otherwise. In addition, the count value in the types of amino acids in which they have appeared should be considered. For example, even if $\mathcal{A}_i = 2$, the meanings of the forms $\{1,4\}$ and $\{2,3\}$ may be different.
To address difference among the count value in each site, we obtained summarized scores for analysis by calculating the weighted mean of scores, with weights based on the count values as follows: 
Define $\mathcal{A}_i=$  ``Types of amino acids appeared at the $i$-th site". Then for the i-th site,
\begin{align*}
& \text{Independent Score} = \left(q_{s_i, s_i} \cdot (1-\frac{n}{8}) \times w_{s_i}\right) \bigg/ \sum_{s_i}w_{s_i}, \\
& \text{Pairwise Score} = \left( q_{s_i, s_j} \cdot \frac{n}{8} \times w'_{s_i, s_j} \right) \bigg/ \sum_{s_i, s_j}w'_{s_i, s_j}, 
\end{align*}
where
\begin{align*}
& s_i \in \mathcal{A}_i, s_i \neq s_j \in \mathcal{A}_i, \\
& n=\#(\mathcal{A}_i), \\
& q_{s_i, s_i}=\text{BLOSUM62}[s_i,s_i],\  q_{s_i, s_j}=\text{BLOSUM62}[s_i,s_j],\\
& w_{s_i}={\bf z_i}[s_i],\  w'_{s_i, s_j}=({\bf z_i}[s_i]+{\bf z_i}[s_j]), \\
& \text{(Note that ${\bf z_i}[s_i]$ indicates $s_i$-th element of ${\bf z_i}.$)}
\end{align*}

\begin{figure}[htp]\centering
\subfloat[]{\label{a}\includegraphics[width=.5\linewidth]{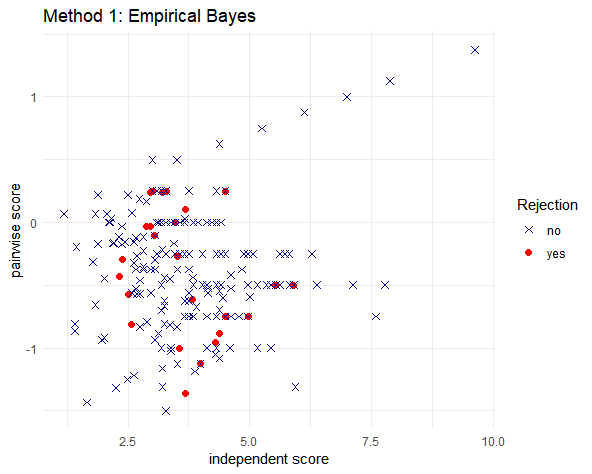}}\par
\subfloat[]{\label{b}\includegraphics[width=.5\linewidth]{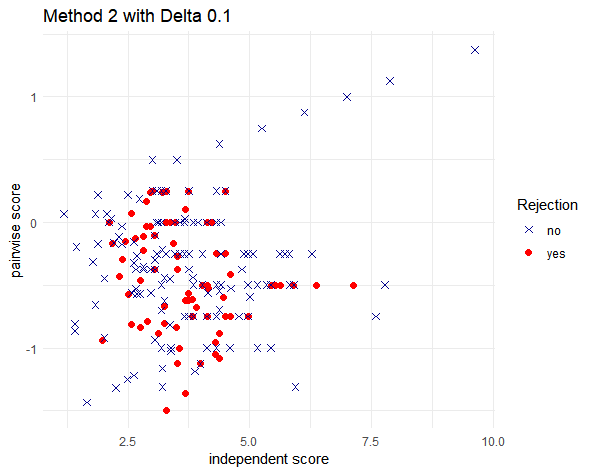}}\hfill 
\subfloat[]{\label{c}\includegraphics[width=.5\linewidth]{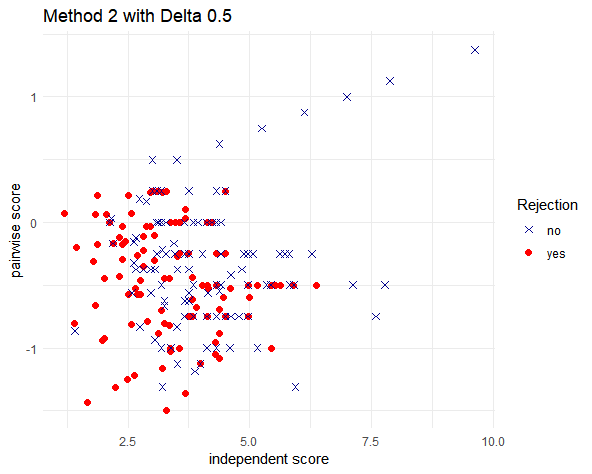}}
\caption{Results of analysis based on independent and pairwise scores for each sites in (a) Model 1, (b) Model 2 when $\delta=0.1$, (c) Model 2 when $\delta=0.5$.}
\label{fig}
\end{figure}

\begin{figure}[htp]
\centering
\subfloat[]{\includegraphics[width=.5\linewidth]{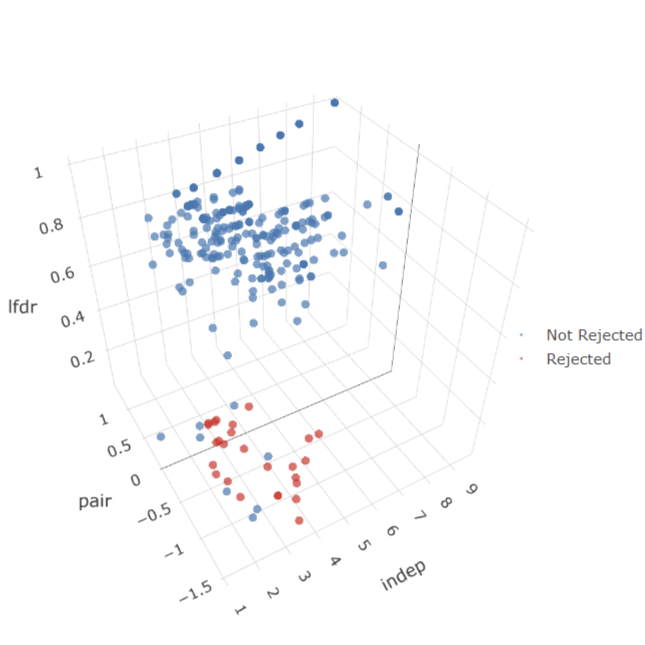}}\par
\subfloat[]{\includegraphics[width=.5\linewidth]{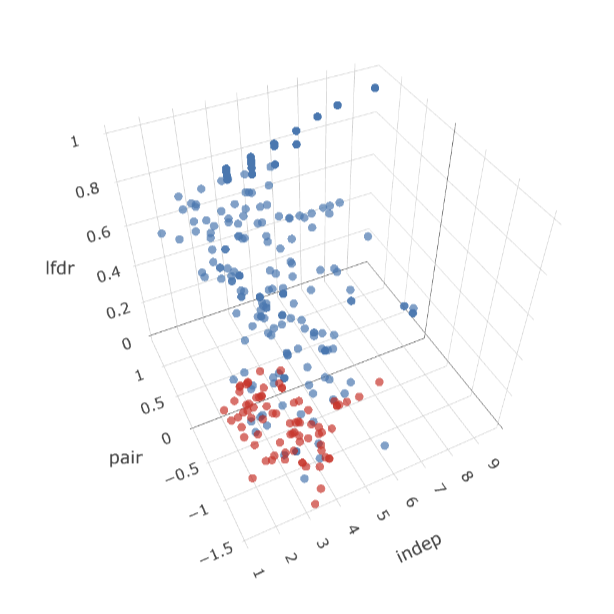}}\hfill 
\subfloat[]{\includegraphics[width=.5\linewidth]{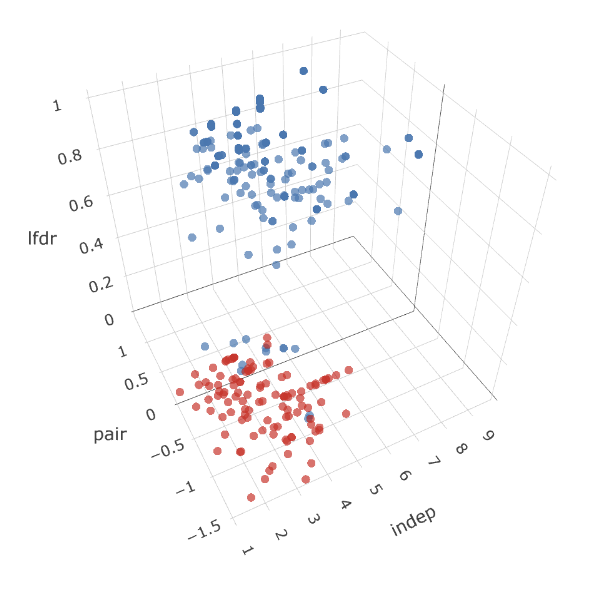}}
\caption{3d Plot (independent score vs pairwise score vs estimated local fdr) for the results of analysis based on (a) Model 1, (b) Model 2 when $\delta=0.1$, (c) Model 2 when $\delta=0.5$.}
\label{fig2}
\end{figure}

Figure 1 shows the results of HIV data analysis based on Model 1 and Model 2 with $\delta=0.1$ and $\delta=0.5$, respectively. Red points in the plots indicate the rejected(detected) sites and Blue points indicate not-rejected sites. Compared to Model 1 which we used empirical Bayes approaches, more rejection sites are observed  in the plots of Model 2 when the scores in $x$, $y$ axis are small, which indicates that the independent and pairwise scores influence the model simultaneously. Also, 3D scatter plots in Figure 2 show the relationship between three variables (independent scores, pairwise scores, and local FDR values) more clearly than in Figure 1, where there was some overlap among points. We observed that more points are rejected towards the bottom-left direction as we incorporate pairwise terms in Model 2. Furthermore, as tuning parameter $\delta$ increases, the impact of the pairwise term within the model becomes more significant. This means that more weight on association of amino acids distinguishes different pairs of amino acids more clearly depending on the scores in the BLOSUM62 substitution matrix. Therefore, the proposed model, which incorporates the pairwise term to give weight to rare events, identified significant sites with lower scores better compared to the model that considered only independent terms.

\section{Conclusion and Discussions}
Sparse count data from two populations arise frequently in HIV virus studies
and one main focus is on identifying the properties that distinguish HIV founder viruses from other circulating viruses in the donor. When data are very sparse in the sense that counts are very low or zeros in many cells, Fisher's exact test is not desirable due to low powers. Also, Benjamini and Hochberg (1995) procedure does not reject any of sites under the nominal false discovery rate of $0.05$. We address the issue of finding statistically significant sites in HIV data since existing traditional approaches does not detect any statistical significant residue. We focused on addressing two points: one is the drawback of Fisher's exact test especially when data are sparse, and the other is the motivation of introducing more biological knowledge into the proposed model to overcome sparsity of data.

Since the occurrence varies among the 20 types of amino acid, it is important to utilize such information to increase the power of detecting significant positions. In this paper, we proposed two models to define False Discovery Rate controlling procedure in identifying significant sites to distinguish Transmitted and Non-Transmitted groups of HIV viruses:  
1) the empirical Bayes model incorporating BLOSUM62 substitution matrix into prior information, 2) Markov random field model which considers pairwise interaction of two amino acids. Both proposed models are based on Bayesian approaches which combines observed data and prior knowledge from the BLOSUM62 substitution matrix. 
The proposed methods identify significant sites while the BH procedure with Fisher's exact test does not discover any significant site as we demonstrated.  
In addition, by simultaneously considering the occurrence of a pair of amino acids from BLOSUM62 substitution matrix, the second approach was able to identify additional significant sites based on scores when those sites have the same constructions in the data.

The main idea of our paper is to incorporate more scientific knowledge into statistical models and derive more reasonable results. Our strategies have the potential to be applied to many data sets from biomedical studies in detecting significant residues in DNA or protein sequences. We expect that our efforts will provide a useful approach for the area of statistics, biology and related fields, especially when handling small sample size datasets while incorporating biological knowledge in statistical models. 

Finally, as part of future work, we suggest further improvement as follows. 
{First, one limitation in our study is that we assumed independence between $P_s$ and $P_{st}$. In realistic, it is possible that these parameters are related to each other. In other words, as the parameter for pairwise term decreases, the parameter for independent term may tend to increase. Although we proceeded sampling under the assumption of independence, we would expect potential improvement if we incorporate their relationships into the model. One possible approach could involve imposing additional restriction on the parameters, such as $\sum_{1\leq s,t \leq 20}p_{st}= \sum_s p_{ss} + 2\sum_{s<t}p_{st} = 1$ and $\sum_{t=1}^{20} p_{st} = p_s$.} 
Second, we pre-specified $\beta_1$ and $\beta_2$ in our Model 2. There may exist other appropriate values for $\beta_1$ and $\beta_2$. Our objective was to generate sample values from posterior so that parameters in Model 2 do not deviate significantly from the expected values based on domain knowledge. Due to the complexity, it was computationally challenging to estimate these parameters through empirical Bayes estimation. We examined some candidates for these parameters and MCMC samples were generated within reasonable bounds when ($\beta_1$, $\beta_2$) is assumed around the values of (1000,10000).  If a more robust method for inferring tuning parameters becomes available, it could lead to more accurate results.
Lastly, we assumed $\delta=0.1, 0.5, 1$ as a tuning parameter over all possible values. The reason for this is that when $\delta$ is close to 0, the meaning of adding a pairwise term diminishes, while when $\delta = 1$, the effect of the pairwise term becomes too strong, offsetting the effect of the independent term and yielding less reasonable results in terms of biological domain. Hence, this tuning parameters can be further improved by utilizing other informative distributions to construct prior.

\section*{Acknowledgement}
This work was supported by the National Research Foundation of Korea (NRF) grant funded by the Korea government (MSIT) (No. 2020R1A2C1A01100526).

{}

\appendix 

\begin{appendices}
\section{Derivation of \texorpdfstring{${q_s}$ 
for {\large $1\leq s \leq 20$} and ${q_{st}}$ for {\large $1\leq s<t\leq 20$.}}{}\label{appendixA}}

In Appendix A, we will explain how the scores in the BLOSUM62 substitution matrix were calculated in \cite{Heinkoff}, then present the process of deriving ${\bf q} =(q_1,\ldots, q_{20}) $ and 
${\bf q}_{assoc} = (q_{1,1}, q_{1,2}, \ldots, q_{19, 20})$ in Model 1 and Model 2 
where ${\bf q}_{assoc}$ represents the vector including associations of $20$ amino acids. 

Let $p_{st}$ be the frequency of occurrence for each $s,t$ amino acid pair and $f_s$ be the background frequency of occurrence for amino acid $s$ in any protein sequence, with constraints $\sum_s f_s=1$, for $1\leq s<t \leq 20$. The Logarithm of Odds value is defined as ${\rm log}_2\left(\frac{p_{st}}{f_sf_t}\right).$
Then, the scores in the BLOSUM62 matrix are calculated by multiplying scaling factor of 2 and rounded to the nearest integer:
$$ {\cal S}_{st} = 2{\rm log}_2\left(\frac{p_{st}}{f_sf_t}\right),$$

Hence, the target frequencies of the BLOSUM62 substitution matrix can be obtained from the corresponding scores.
\renewcommand\labelitemi{{\boldmath$\cdot$}}
\begin{itemize}
    \item From the definition of BLOSUM62, we know $2^{\frac{{\cal S}_{st}}{2}} = \frac{p_{st}}{f_sf_t}.$
    \item Obtain target frequencies $p_{st} = f_s \cdot f_t \cdot 2^{{{\cal S}_{st}}/{2}}.$

\end{itemize}
Finally, we accordingly derive ${\bf q}= (q_1,...,q_{20})$ and ${\bf q}_{assoc}=(q_{1,2},...,q_{19,20})$ for the model 1 and 2 using these frequencies: 
\begin{align*}
    & q_s = \sum_{t}p_{st}, \\
    & q_{st} = \frac{p_{st}}{\sum_{s<t}p_{st}}, \text{for $1\leq s < t \leq 20$.}
\end{align*}

As mentioned in section 4.2.1, we assumed independence between ${\bf q}$ and ${\bf q}_{assoc}$ with constraint $\sum_s q_s =1$ and $\sum_{s \neq t}q_{st}= 2 \sum_{s<t} q_{st} = 1.$ 
These derived values reflect the size of the BLOSUM62 matrix score and provide weights based on whether the pair is rare or common in our proposed model. The background frequencies $f_s$ for the BLOSUM62 substitution matrix were obtained from \url{https://www.ncbi.nlm.nih.gov/IEB/ToolBox/CPP_DOC/lxr/source/src/algo/blast/composition_adjustment/matrix_frequency_data.c} in \cite{blosum}.

\end{appendices}

\end{document}